\documentclass[aps,twocolumn,prl,showpacs]{revtex4}

\usepackage{graphicx}
\usepackage{dcolumn}
\usepackage{bm}

\begin{document}

\preprint{physics/0602067}

\title{Geometrical Dynamics in a Transitioning Superconducting Sphere}

\author{James R. Claycomb}
\email{jclaycomb@hbu.edu}
\altaffiliation[also at ]{T$_C$SUH, University of Houston}
\affiliation{Department of Mathematics and Physics \\
Houston Baptist University \\
Houston, TX 77074-3298, USA}%

\author{Rambis K. Chu}
\email{chu_rk@tsu.edu}
\affiliation{Bio-Nano Computational Laboratory RCMI-NCRR \\
Physics Department \\
Texas Southern University \\
Houston, TX 77004, USA}%

\date{\today}

\begin{abstract}
Recent theoretical works have concentrated on calculating the Casimir effect in curved spacetime.
In this paper we outline the forward problem of metrical variation due to the Casimir effect for spherical geometries.
We consider a scalar quantum field inside a hollow superconducting sphere.
Metric equations are developed describing the evolution of the scalar curvature
after the sphere transitions to the normal state.
\end{abstract}


\maketitle

\section{INTRODUCTION}

The classical Casimir effect \cite{cas48,mil94} may be viewed as vacuum reduction by mode truncation
where the presence of conducting boundaries, or capacitor plates, excludes vacuum modes with wavelengths
longer than the separation between the conductors. The exclusion of longer wavelengths
results in a lower vacuum pressure between the plates than in external regions. The resulting pressure difference,
or Casimir force, may act to push the conductors together, effectively collapsing the reduced vacuum phase.
This tiny force has been measured experimentally \cite{lam97,moh98} in agreement with the predictions of quantum electrodynamics. 
Boyer gives the first detailed treatment of the vacuum modes inside a conducting sphere \cite{boy68}
with more a recent account by Milton \cite{mil97}. The Casimir effect for spherical conducting shells
in external electromagnetic fields has been investigated \cite{mil78,nes98}. Applications of the Casimir effect
to the bag model have been studied for massive scalar \cite{bor97} and Dirac \cite{eli98} fields
confined to the interior of the shell. An example of the Casimir effect in curved spacetime
has been considered for spherical geometries \cite{bay98} in de Sitter space \cite{set011}
and in the background of static domain wall \cite{set012}. In this paper we investigate the metrical variations
resulting from vacuum pressure differences established by a spherical superconducting boundary.
We first consider the static case when the sphere is superconducting and then the dynamical case
as the sphere passes to the normal state. 

\section{THE STATIC CASE}

Our idealized massless, thin sphere of radius $R_{0}$ has zero conductivity in the normal state.
In the superconducting state, the vacuum inside the hollow is reduced so that there exists a pressure difference
$\Delta p$ inside and outside the sphere. In general, all quantum fields will contribute to the vacuum energy.
When the sphere of volume $V$ transitions to the superconducting state, a latent heat of vacuum phase transition
$\Delta pV$ is exchanged. The distribution of vacuum pressure, energy density and space-time geometry are described by
the semi-classical Einstein field equations taking $c$ = 1,

\begin{equation}
R_{\mu\nu} - \frac{1}{2}Rg_{\mu\nu} = 8\pi G \left\langle T_{\mu\nu} \right\rangle
\end{equation}

where $R_{\mu\nu}$ and $R$ are the Ricci tensor and scalar curvature, respectively. $\left\langle T_{\mu\nu} \right\rangle$
is the vacuum expectation of the stress energy tensor. Regulation procedures for calculating the renormalized
stress energy tensor are given in \cite{bir94} for various geometries. The most general metric with spherical symmetry 

\begin{eqnarray}
ds^{2} = B\left(r,t\right) dt^{2} - A\left(r,t\right) dr^{2} - C\left(r,t\right) dr dt \nonumber\\
- r^{2} d\theta^{2} - r^{2} sin^{2}\theta d\phi^{2}
\end{eqnarray}

where $A$, $B$, and $C$ are arbitrary functions of time and the radial coordinate. (2) can be written
under normal coordinate transformation \cite{wei72},

\begin{eqnarray}
ds^{2} = \tilde{B}\left(r,t\right) dt^{2} - \tilde{A}\left(r,t\right) dr^{2} 
- r^{2} d\theta^{2} - r^{2} sin^{2}\theta d\phi^{2}
\end{eqnarray}

The metric tensor then becomes, dropping tildes,

\begin{equation}
g_{\mu\nu} = Diag \left( B\left(r,t\right), A\left(r,t\right), -r^{2}, - r^{2} sin^{2} \right)
\end{equation}

For a diagonal stress energy tensor, the solutions to equation (3) relating $A$ and $B$ are

\begin{equation}
- \frac{1}{r^{2}} + \frac{1}{r^{2}A} - \frac{A'}{rA^{2}} = 8\pi G \frac{1}{B}\left\langle T_{00} \right\rangle
\end{equation}

\begin{equation}
\frac{1}{r^{2}} - \frac{1}{r^{2}A} - \frac{B'}{rAB} = 8\pi G \frac{1}{A}\left\langle T_{11} \right\rangle
\end{equation}

\begin{eqnarray}
\frac{A'}{2rA^{2}} - \frac{B'}{2rAB} + \frac{A'B'}{4A^{2}B} - \frac{B'^{2}}{4AB^{2}} \nonumber\\
- \frac{B''}{2AB} = 8\pi G \frac{1}{r^{2}}\left\langle T_{22} \right\rangle
\end{eqnarray}

with a fourth equation identical to (7). The prime denotes  $\partial_{r}$.
Note that all time derivatives cancel from the field equations when the metric is in standard form
and the stress energy tensor is diagonal. When the sphere is in the superconducting state,
the scalar curvature $R=g^{\mu\nu}R_{\mu\nu}$ is given by

\begin{eqnarray}
R = \frac{2}{r^{2}} - \frac{2}{r^{2}A} + \frac{2A'}{rA^{2}} - \frac{2B'}{rAB} \nonumber\\
+ \frac{A'B'}{2A^{2}B} + \frac{B'^{2}}{2AB^{2}} - \frac{B''}{AB}
\end{eqnarray}

In calculating the Casimir force, one properly calculates differences in vacuum pressure
established by the conducting boundaries \cite{mil94}. In the present case,
it is only meaningful to consider changes in scalar curvature due to variations in vacuum pressure. 

\section{THE DYNAMICAL CASE}

If the sphere passes from the superconducting to the normal state, the pressure should equalize as the vacuum relaxes.
The diagonal form of the stress energy tensor results in the cancellation of all time derivatives in the field equations.
External electromagnetic fields will contribute off-diagonal terms, however we wish to consider
how the pressure equalizes in absence of external fields. The key is that the required time dependence
is provided by the zero point field fluctuations. As the simplest example, we consider the massless scalar quantum field
with stress energy tensor \cite{bir94}.     

\begin{equation}
T_{\mu\nu} = \phi_{,\mu}\phi_{,\mu} - \frac{1}{2} g_{\mu\nu} g^{\alpha\beta} \phi_{,\alpha}\phi_{,\beta}
\end{equation}

The non-zero components of T are 

\begin{equation}
T_{00} = \frac{1}{2} \dot{\phi^{2}} + \frac{B}{2A} \phi'^{2}
\end{equation}

\begin{equation}
T_{11} = \frac{1}{2} \phi'^{2} + \frac{A}{2B} \dot{\phi}^{2}
\end{equation}

\begin{equation}
T_{22} = r^{2} \left( \frac{1}{2B} \dot{\phi}^{2} + \frac{1}{2A} \phi'^{2} \right)
\end{equation}

\begin{equation}
T_{33} = r^{2} sin^{2} \theta \left( \frac{1}{2B} \dot{\phi}^{2} + \frac{1}{2A} \phi'^{2} \right)
\end{equation}

\begin{equation}
T_{01} = \dot{\phi} \phi'
\end{equation}

where $T_{01}$ = $T_{10}$. The semi-classical field equations become

\begin{equation}
- \frac{1}{r^{2}} + \frac{1}{r^{2}A} - \frac{A'}{rA^{2}} = 8\pi G \frac{1}{B}\left\langle T_{00} \right\rangle
\end{equation}

\begin{equation}
\frac{1}{r^{2}} - \frac{1}{r^{2}A} - \frac{B'}{rAB} = 8\pi G \frac{1}{A}\left\langle T_{11} \right\rangle
\end{equation}

\begin{eqnarray}
- \frac{\dot{A}^{2}}{4A^{2}B} - \frac{\dot{A}\dot{B}}{4AB^{2}} + \frac{\ddot{A}}{2AB} + \frac{A'}{2r^{2}A} \nonumber\\
- \frac{B'}{2rAB} + \frac{A'B'}{4A^{2}B} + \frac{B'^{2}}{4AB^{2}} - \frac{B''}{2AB} \nonumber\\
= 8\pi G \frac{1}{r^{2}}\left\langle T_{22} \right\rangle
\end{eqnarray}

\begin{equation}
- \frac{\dot{A}}{rA} = 8\pi G \left\langle T_{01} \right\rangle
\end{equation}

Equations (15) and (16) are identical to (5) and (6). Two additional equations are identical to (17) and (18).
Expressions for $A$ and $B$ may be obtained from equation (18) and (15) or (16), respectively.
The scalar curvature is given by

\begin{eqnarray}
R = \frac{2}{r^{2}} - \frac{2}{r^{2}A} - \frac{\dot{A}^{2}}{2A^{2}B} - \frac{\dot{A}\dot{B}}{AB} \nonumber\\
+ \frac{\ddot{A}}{AB} + \frac{2A'}{rA^{2}} - \frac{2B'}{rAB} + \frac{A'B'}{2A^{2}B} \nonumber\\
+ \frac{B'^{2}}{2AB^{2}} - \frac{B''}{AB}
\end{eqnarray}

Combining equation (19) with (15-17) and (10-12) reveals

\begin{equation}
R = 16\pi G \left\langle \frac{\dot{\phi}^{2}}{2B} - \frac{\phi'^{2}}{2A} \right\rangle
\end{equation}

When evaluating changes in scalar curvature, the expression for $R$ in absence of the sphere should be subtracted
from that obtained for a given quantum field. 

\section{CONCLUSION}

When a hollow sphere transitions between the normal and superconducting state a latent heat
of vacuum phase transition is exchanged. In the dynamical case, zero-point field fluctuations result
in off-diagonal components of the stress energy tensor that give rise to time dependent field equations.
The analysis presented here may be extended to include massive fields with coupling or spin
(0, $\frac{1}{2}$ and 1) as well as other superconducting geometries. 

\section{ACKNOWLEDGMENTS}
This work was supported, in part, by RCMI through NCRR-NIH (Grant No. G12-RR-03045),
Texas Southern University Research Seed Grant 2004/2005, and the State of Texas
through the Texas Center for Superconductivity at University of Houston.

\bibliography{cas_hts}

\end{document}